\documentclass[a4paper, amsfonts, amssymb, amsmath, reprint, showkeys, nofootinbib, twoside]{revtex4-2}
\usepackage[english]{babel}
\usepackage[utf8]{inputenc}
\usepackage{xcolor}
\usepackage{lipsum}
\usepackage[colorinlistoftodos, color=green!40, prependcaption]{todonotes}

\usepackage[pdftex, pdftitle={Article}, pdfauthor={Author}]{hyperref} % For hyperlinks in the PDF
\newcommand{\makefigure}[2]{
\begin{figure}[ht!]
\begin{center}
\includegraphics[width=0.95\columnwidth]{#1}
\caption{#2 \label{fig:#1}}
\end{center}
\end{figure}}

\graphicspath{{images/}}

\begin{document}

\title{Size Effect on Raman Measured Stress and Strain Induced Phonon Shifts in Ultra-Thin Film Silicon}

\author{C.~Pashartis}
\email[Correspondence email address: ]{christopher.pashartis@imec.be}
\affiliation{IMEC, Leuven, Belgium}
\author{M.~J.~van~Setten}
\affiliation{IMEC, Leuven, Belgium}
\affiliation{ETSF, European Theoretical Spectroscopy Facility}
\author{G.~Pourtois}
\affiliation{IMEC, Leuven, Belgium}
\date{\today}

% %-----------------------------------------------------------------------
% %
% %                       A B S T R A C T
% %
% %-----------------------------------------------------------------------

\begin{abstract}
%Raman spectroscopy relies on knowing the mechanical properties of the material \textit{a priori} to measure the stress in a material. As semiconductor device features are now on the order of less than 10~nm, we demonstrate that for the same Raman-active phonon frequency shifts observed in a bulk material, significantly more stress must be appropriately measured. In order to use Raman spectroscopy to properly measure changes in stress at the interface or in thin layers of a semiconductor device, the correct elastic tensor and phonon deformation potentials must be used and compiled by the scientific community.

The fabrication of complex nano-scale structures, which is a crucial step in the scaling of (nano) electronic devices, often leads to residual stress in the different layers present. This stress gradient can change many of the material properties and leads to desired or undesired effects, especially in the active part of the transistor, its channel. Measuring, understanding, and, ultimately, controlling the stress fields is hence crucial for many design steps.The level of stress can in principle be measured by micro-Raman spectroscopy. This, however, requires \emph{a priori} knowledge of the mechanical properties of the material. The mechanical properties start to deviate from the bulk values when film dimensions become thinner than 5 nm. If this effect is ignored, errors of up to 400\% can be introduced in the extracted stress profile. In this work, we illustrate this effect for a range of Si (001) slabs with different silicon film thickness, ranging from 5 to 0.7 nm and provide best practices for the proper interpretation of micro-Raman stress measurements.

\end{abstract}

% %-----------------------------------------------------------------------
% %
% %                       A B S T R A C T
% %
% %-----------------------------------------------------------------------


\maketitle

%%%%%%%%%%%% Focus 3750 words for PRL
% rename sections?
% \section{Introduction}

% \section{Methodology}

% \section{Results \& Discussion}

% \section{Conclusion}

%%%%%%%%%%%%%%%%%%%%%%%%%%%%%%%%%%
%importance of raman spectroscopy
%%%%%%%%%%%%%%%%%%%%%%%%%%%%%%%%%%
Raman spectroscopy is a key characterization technique for measuring vibrational modes in biology, chemistry, and material science. The technique has proven to be an excellent non-invasive source of information in medicine, where it can identify protein morphologies and their structural evolution.\cite{Schulz_VS_2007, Wen_JOPS_2007, Shashilov_JORSAIJFOWIAAORSIHOPAABARS_2009, Rygula_JORS_2013} In chemistry, surface-enhanced Raman scattering is a capable variant, which is able to detect vibrational signatures from single molecules, usually identified by fluorescence.\cite{Nie_S_1997, Kneipp_PRL_1997} In applications in solid state and semiconductor physics, Raman spectroscopy is also often used to determine the amount of stress engineered / present in the active component of a transistor. Using Raman peak shifts to determine stress, however, requires the a priory knowledge of the elastic \cite{DeWolf_SSAT_1996, DeWolf_JOAP_1996} or the Grüneisen tensor\cite{Angel_ZFKM_2019}. The wavelength of light gives an indication of penetration depth of the material, for example, infrared light can penetrate to a few $\mu$m and at the UV wavelength it can be a few nm \cite{Born__2013}. Using deep penetration within a surface, stress fields in different layers of nano structures can be measured with the additional advantage of being non-destructive. For reduced dimension, these start to deviate from their bulk values as quantified for free-standing ultra-thin films, observed in 2D systems,\cite{Lee_S_2008, Bertolazzi_AN_2011} nano membranes,\cite{Sarkar_NR_2021} and nanowires \cite{Lee_PRB_2007}. 

Since modern devices in nano electronics contain many material layers and have feature sizes on the order of 2-10~nm, interfacial or surface stresses become increasingly important. This is especially the case if the materials become less stiff in the confinement direction(s), i.e. yielding higher amounts of strain for the same unit of stress, under thinning. Since the conversion of the shifts in Raman frequencies into a stress tensor relies on the actual knowledge of the material elastic tensor, the question arises: without the quantification of the proper thickness dependence of the tensor, how much is the technique exaggerating the measured stress? In this letter, we address the problem in two ways. Firstly, by using existing knowledge of the elastic tensor variations with thinning. Secondly, we compute the shift of the peak phonon frequencies under external strain, using \textit{ab initio} Si thin films models in the [001] direction with \texttt{ABINIT} \cite{Gonze_CPC_2020, Romero_JCP_2020}. Therefore, it should be no surprise that there is also a size effect on the phonon frequencies of crystalline materials, yielding changing Raman shifts or broadening, which in tiny systems can be attributed to surface frequencies.\cite{Wang_MCAP_2001, Choi_VS_2005, Werninghaus_APL_1997}

\subsection{Overestimation Error Approximation}
We begin our analysis by assessing the amount of overestimation by first estimating the extent of the effect by calculating the phonon shift with incident light polarized perpendicular to the surface on a film strained in the same direction (see \autoref{sec:appendix}):
\begin{equation} \label{eqn:constitutive_equation_Si_001_33}
    \Delta \omega_3 = \frac{1}{2\omega_0} \left[(p S_{33} + q (S_{31}+S_{32}) \right]\sigma_{33}, \hspace{.5cm} \lambda = \Delta \omega \cdot 2 \omega_0 \hspace{0.1cm}.
\end{equation}
Here, we implicitly assume that the ultra-thin film has the same three-fold degenerate phonons as the primitive Si system (or 3 Raman active modes)  and that the phonon deformation constants are the same as those of the bulk. We define $\Delta \omega_3$ as the third phonon frequency dependent on the $\sigma_{33}$ Cauchy-stress component (the one in the direction of the thickness of the film), and $\omega_0$ as the non-perturbed phonon frequency. We report the Hartree-corrected elastic components for 1-5~nm Si films oriented in the [001] direction with dimer reconstruction in \autoref{tab:silicon_c_values}, see \autoref{sec:appendix} for details. For a 1~nm Si film oriented with the vacuum in the Cartesian-$\hat{z}$ direction, the $c_{33}$ ($zz$) component reduces from 155~GPa in bulk to 51~GPa. Correspondingly, at 3~nm, the elastic tensor component was found to be 103~GPa and at 5~nm, it further increases to 121~GPa. Bulk Si has a peak phonon frequency of 520.5~cm$^{-1}$ (observed at the $\Gamma$-symmetry point in the Brillouin zone), with Phonon Deformation Potentials (PDPs) of $p=-1.85\cdot w_0^2$ and $q=-1.85\cdot w_0^2$.\cite{Cerdeira_PRB_1972,Anastassakis_PRB_1990, Tuschel_S_2019} In \autoref{fig:001_light_001_stress_raman_hartree_corrected}, it is observed that for the same magnitude of change in peak Raman frequency observed for a material under strain ($\Delta \omega$), the DFT-computed free standing films yield far less stress than the bulk crystalline system.

% table of Si data
\begin{table}
% \centering
% \resizebox{0.5\textwidth}{!}{%
\begin{tabular}{c|c c c c c c c c c c c c}
\hline 
System & $c_{11}$ & $c_{12}$ & $c_{13}$ & $c_{21}$ & $c_{22}$ & $c_{23}$ & $c_{31}$ & $c_{32}$ & $c_{33}$ & $c_{44}$ & $c_{55}$ & $c_{66}$ \\
\hline
Bulk & 155 & 114 & 114 & 114 & 155 & 114 & 114 & 114 & 155 & 77 & 77 & 77 \\
51~\AA & 150 & 119 & 107 & 119 & 150 & 107 & 107 & 107 & 121 & 66 & 66 & 73 \\
29~\AA & 145 & 122 & 102 & 122 & 145 & 103 & 102 & 103 & 103 & 59 & 58 & 69 \\
7~\AA & 103 & 100 & 82 & 100 & 162 & 86 & 82 & 86 & 51 & 38 & 35 & 46 \\
\hline
\hline
\end{tabular}%
% }
\caption{Elastic tensor components in GPa used to compute the stiffness tensor in the initial approximation of \autoref{fig:001_light_001_stress_raman_hartree_corrected}.}
\label{tab:silicon_c_values}
\end{table}

% calculation under assumption
As a relatively quick proof of concept, the experimental values of the bulk elastic tensor from Ref.~\cite{Anastassakis_PRB_1990} are used to compute the slopes and show good agreement with the computed crystalline system (blue line). The evaluation of the slopes of changing Raman shifts with externally applied stress demonstrates that if the PDPs do not change significantly, the stress in ultra-thin films in silicon device architecture is vastly overestimated. At its worst, the 1~nm equivalent film with a shift of 0.2~cm$^{-1}$ has a stress of 33~MPa, whereas if the bulk compliance tensor was used on the same system, the corresponding stress is over 140~MPa - a factor of more than four increase. Even at a thickness of 5~nm, we obtain a corrected stress of 108~MPa, giving an error of over 30~MPa if the bulk parameters were used. Once again, the assumption takes for granted that the symmetry of the elastic tensor and PDP tensor remain equivalent in the bulk and in the free-standing film - which is evidently not the case. Regardless, this simple example neatly outlines how drastic the consequences of assuming the bulk properties when trying to quantify stresses in ultra-thin film materials may be; suggesting that a more in-depth approach to verify the variation in phonon frequencies corresponding to the Raman shifts is warranted.

\makefigure{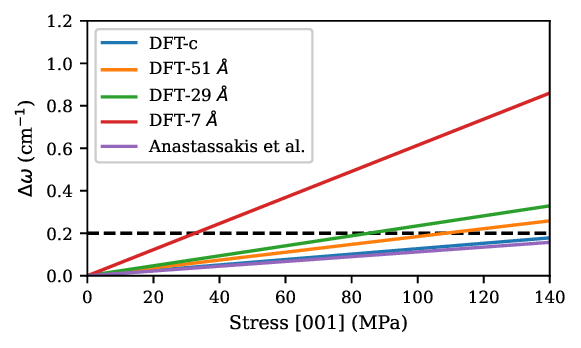}{Calculations for the Raman shift of silicon assuming the phonon deformation potentials remain unchanged and only the elastic tensor varies. Bulk and free-standing reconstructed Si $2\times1$ (001) surfaces of different thicknesses are given (-c corresponds to crystalline bulk, thin films denoted by their thicknesses). Experimental crystalline Ref.~\cite{Anastassakis_PRB_1990} is given for reference. The dotted black line is a guide for measuring the same phonon frequency shift.}

%%%%%%%%%%%%%%%%%%%%%%%%%%%%%%%%%%
% so how do we go about doing this wth DFT?
%%%%%%%%%%%%%%%%%%%%%%%%%%%%%%%%%%
% compute bulk phonon example
\subsection{Bulk Phonon Shifts Under Strain}
Instead of computing the elastic tensor and the phonon deformation potentials, the remaining portion of this letter focuses on directly calculating the phonon frequency shifts corresponding to Raman active modes, under externally applied strain. We first demonstrate that the correct phonon frequencies can be computed for cubic crystalline silicon. In the primitive unit cell of Si, consisting of two atoms, one would expect six phonon frequencies, three of which are the acoustic modes at 0~cm$^{-1}$ with the other three being the optical frequencies near 520.5~cm$^{-1}$.\cite{Cerdeira_PRB_1972} Existing Density Functional Perturbation Theory (DFPT) calculations are in good agreement at 510.7~cm$^{-1}$ using the PBEsol exchange correlation in  \texttt{Abinit}.\cite{Petretto_SD_2018} Our calculations show good agreement with a peak average $\Gamma$ frequency of 506.4~cm$^{-1}$, being about 3\% smaller than the experimental value. However, we are primarily interested in the change in phonon frequencies corresponding to the shift in Raman peak frequencies in the Si (001) $2\times1$ surface.

% To ensure DFT optimized parameter transferability to the Si (001) $2\times1$ surface, a cubic unit cell was used, resulting in the phonon modes in \autoref{fig:6.1:phonon_freq_cubic_0}.
Compared to the 2-atom primitive cell of Si, since the cubic cell is not equivalent to the primitive cell, other Brillouin Zone (BZ) symmetry points are folded onto the $\Gamma$-point giving the intermediate modes in between the expected modes. In the same manner, it is expected that some BZ points will wrap onto the $\Gamma$-point in the surface model, even though its smallest unit of periodicity is the $2\times1$ surface reconstruction. The eigenvectors and vibrational movement of atomic positions in bulk Si correspond to nearest atoms oscillating out-of-phase. This effectively creates a dipole oscillation resulting in a longitudinal optical phonon mode, the three maxima frequencies. These optical modes are those that are visible by the Raman scattering technique.

% now do strain
In order to compare to the exact same conditions of \autoref{fig:001_light_001_stress_raman_hartree_corrected}, we are most concerned with the change in Raman peak frequencies in the same direction. This can be achieved by applying strain to the cubic cell along the [001] direction while the other lattice directions of [100] and [010] are allowed to contract and expand in order to minimize the stress in directions other than $\sigma_{33}$. Successive tensile strains ranging from 0 to 2\% were applied to the system, as it tends to numerically perform in a more stable manner than compressive strains. The resulting stress from the material body can then be used in conjunction with the phonon analysis to reproduce the expected slope of \autoref{fig:001_light_001_stress_raman_hartree_corrected} for bulk crystalline Si. \autoref{fig:phonon_freq_bulk_8_atom_peak_strain_stress_single} demonstrates the accuracy of the \textit{ab-initio} calculation to detect shifts in the phonon frequency comparable with the experimental results. The inset shows the frequency shift as a function of the stress, in the same range as the earlier calculation with the elastic tensors. The change in phonon frequency (or Raman peak frequency) with the stress applied corresponds to 0.001142~cm$^{-1}$MPa$^{-1}$, which is in agreement with the experimental slope of the equivalent system setup of 0.001124~cm$^{-1}$MPa$^{-1}$ \cite{Anastassakis_PRB_1990}. Reproducing the slope to within 1.6\% difference suggests that using similar convergence parameters on a Si surface yields accurate results.

\makefigure{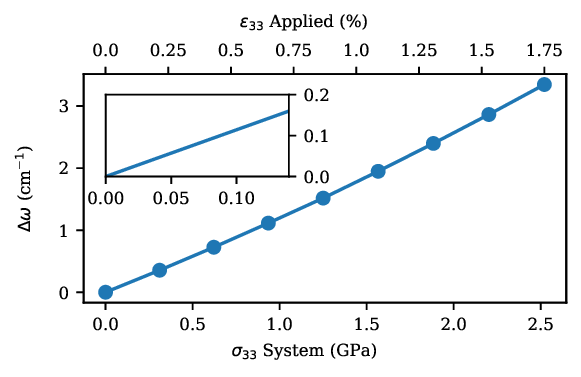}{Shift of averaged phonon frequencies as a function of the stress (bottom axis) with the compressive strain applied (top axis) in the bulk Si system. Data was acquired with tensile strain and transformed to be compressive (pictured) under the assumption of the symmetric elastic regime. Inset figure to be compared with \autoref{fig:001_light_001_stress_raman_hartree_corrected}.}

%%%%%%%%%%%%%%%%%%%%%%%%%%%%%%%%%%
% no we go to the film
\subsection{Ultra-Thin Film Phonon Shifts Under Strain}
To demonstrate the initial hypothesis that the phonon frequency shifts cannot be treated as bulk-like in the regime of ultra-thin films, frequencies of a clean free-standing slab of approximately 1~nm in thickness (i.e. 8 atomic layers) were calculated with the $2\times1$ reconstruction surface in the (001) plane. These 32 atoms yield a total of 96 individual phonon frequencies, as observed in \autoref{fig:modes_1nm_long}. A single peak phonon frequency coincides with the bulk system. The increased number of modes is problematic since the optical modes must be distinguished to determine which phonon frequencies are considered to be Raman-active. In bulk Si the peak frequency is the only Raman active mode, however, in the ultra-thin film are more modes that are potentially active due to symmetry reduction and additional unique atomic sites. We assume in this paper that the predominant Raman active modes are those that are derived from the bulk peak, i.e., should yield a frequency close to the bulk $\Gamma$-point. Therefore, we analyze the Raman modes within 20~cm$^{-1}$ for the 1~nm system. Note that for the following discussion, the first mode of the system is denoted as the zeroth mode.

\makefigure{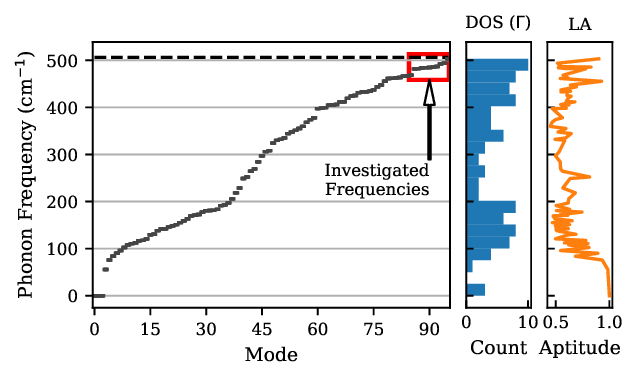}{Phonon frequencies of a 1~nm (8 atomic layered) Si $2\times1$ (001) slab, the dotted black line signifying the peak mode of the bulk Si system, followed by the density of states (DOS) at the $\Gamma$-point (from the modes), and the longitudinal aptitude (LA), respectively. The frequencies studied for Raman activity are denoted by the red box.}

%%%%%%%%%%%%%%%%%%%%%%%%%%%%%%%%%%
% the modes and argument of what we use
A quick analysis of the eigenvectors was performed to determine how longitudinal each mode is by taking the sum of the absolute value of inner product of each vector with every other vector (what we refer to in \autoref{fig:modes_1nm_long} as LA). On average, the closer the sum is to unity, the more longitudinal the character of the phonon frequency is. By applying this scheme, the most energetic mode is longitudinal with a value of 0.90. This was verified by noticing that the eigenvectors or displacements of the phonons in \autoref{fig:1nm_phonon_movement}a) are predominantly bulk-like in motion. Recall that in the bulk, nearest neighbours oscillate forwards and backwards from each other, producing a LO phonon frequency. Typically, we observe that as the eigenfrequency becomes less, relative to the bulk-like mode, the system begins to prefer more movement near the surface. In \autoref{fig:1nm_phonon_movement}b), the next largest frequency is without doubt, not longitudinal and is less ordered. Typically, to determine if a mode is Raman-active, a dipole moment would be associated with the displacement pattern.\cite{Weber__2000} Alternatively, the Raman intensities can be directly computed \textit{ab-initio} by calculating a few parameters such as the dielectric function.\cite{Gillet_PRB_2013} The inclusion of vacuum in the unit cell complicates the Raman intensity calculation since the vacuum height scales the dielectric function, giving erroneous results. It can also quickly become a memory intensive application, so we choose not to compute the intensity here. \textbf{[ref to the code we used to analyze, see materialsproject]}

\makefigure{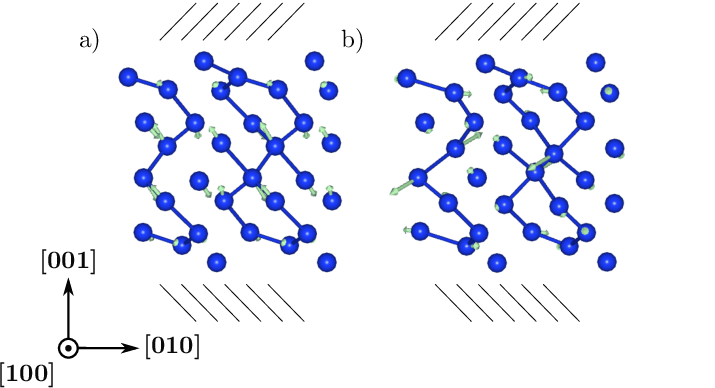}{The two highest energy modes in a Si $2\times1$ (001) slab corresponding to 504 and 495~cm$^{-1}$, respectively. Arrows designate the direction of atomic motion of the phonon frequency. Hashed lines designate the vacuum.}

The modes within 20~cm$^{-1}$ of the most energetic one are given in \autoref{fig:top_20_cm_modes} as a function of strain. With exception of the 93-rd mode, the phonon frequency shift is positive for a unit of compressive strain. This particular mode is characterized by a significant motion between the two layers of central atoms in the slab and may be due to a non-gamma centred mode wrapping onto the $\Gamma$-point, though its displacement is similar to mode 94 which is slightly higher in energy by 4.5~cm$^{-1}$. Given that the motion is similar to a higher energy mode (which points to a numerical issue), we choose to neglect it when calculating the mean phonon shift. \autoref{fig:phonon_slab_85_88_single} shows the results of averaging the shift in phonon frequencies for two regimes: i) including the modes above the 88-th and ii) those above the 85-th, inclusively. The inset is centred in the same region as \autoref{fig:phonon_freq_bulk_8_atom_peak_strain_stress_single} and verifies that the slope does increase; suggesting that for the same unit of stress applied to the system, a higher Raman peak frequency shift should be expected. For the 88-th mode analysis, the slope between the first two points is 0.006310~cm$^{-1}$MPa$^{-1}$ whereas since the 85-th mode includes higher shifts, the slope increases to 0.00775~cm$^{-1}$MPa$^{-1}$. When compared to the bulk crystalline expected relation of 0.001124~cm$^{-1}$MPa$^{-1}$, the slope is nearly 6 times the value due to the change introduced by the surface relative to the bulk centre of the film. Note that this may be an overestimation if the optical modes studied are not $\Gamma$ centred but instead wrapped from the Brillouin-zone, and should therefore not be included in the average top-most modes. Given that 32-atoms are required in the unit cell to capture the surface reconstruction effect, we believe these states to not be wrapped from the Brillouin-zone but being instead, an unique signature of the unit cell. Regardless, even considering just the highest eigenfrequency mode shifts, there is a significant difference in the shift/stress relation compared to the bulk.

\makefigure{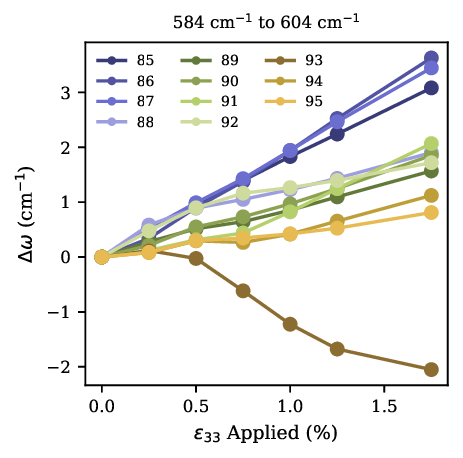}{The modes falling within 20~cm$^{-1}$ of the highest energy eigenfrequency, the reference mode being the 0-th mode. Data was acquired with tensile strain and transformed to be compressive (pictured) under the assumption of the symmetric elastic regime.}

\makefigure{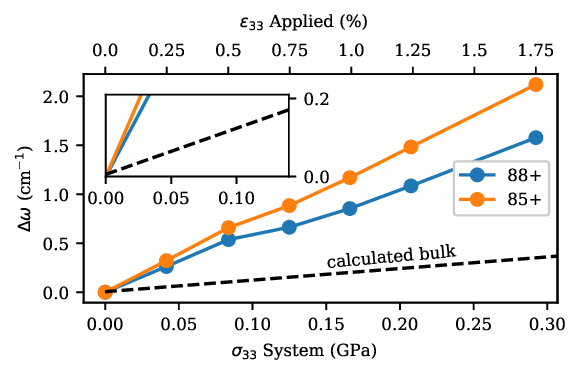}{Shift of averaged phonon frequencies as a function of the resultant stress (bottom axis) with the compressive strain applied (top axis) in Si (001) free-standing slabs. The blue lines entail all modes above the 88th, whereas the orange has those above the 85th, inclusive. The dotted line is the reference line calculated from bulk Si phonon calculations. Data was acquired with tensile strain and transformed to be compressive (pictured) under the assumption of the symmetric elastic regime. Inset figure to be compared with \autoref{fig:001_light_001_stress_raman_hartree_corrected}.}

\subsection{Best Practices}
In order to use Raman spectroscopy effectively in material layers of $<$~5~nm to measure stress, the elastic tensor must be re-evaluated first. Should there be significant interfaces of incredibly thin layers, expect the result to change even more.

\subsection{Summary}
In this letter, we studied Raman peak shifts of silicon ranging from bulk to ultra-thin films using \textit{ab initio} computations. We have shown that experimental PDPs combined with the elastic tensors of ultra-thin freestanding films calculated from Density Functional Theory, suggest that using Raman spectroscopy is vastly overestimating its characterization of the stress. We verified the calculation for bulk Si using \textit{ab-initio} simulations to demonstrate that the phonon frequency shifts do indeed coincide with the expected Raman peak shifts under strain. Finally, by performing the same calculation as the bulk on free-standing ultra-thin film, we expect that indeed the Raman frequency shifts in a 1~nm thick layer can be up to 6 times larger for the same unit of applied stress or strain on the system. For packaged semiconductor devices, this result suggests that material layers are far more strained for significantly less stress. Assuming the bulk elastic tensor values for this characterization greatly overestimates the amount of stress.

\section{Acknowledgements}

The various members of our group for discussions and aid over the course of this research. In particular, Ingrid De Wolf gave many valuable suggestions as to how to properly frame this work.

\bibliography{Main.bib}

\clearpage

\appendix\label{sec:appendix}
\section{Appendix: Cubic Symmetry and Stress Characterization}
The original derivation for what follows is modelled after unixial stress applied in the [110] direction by De Wolf 2015 \cite{DeWolf_JOAP_2015}, but is later extended to an orthorhombic system with strain (or its equivalent stress) in the [001] direction. For a system with diamond cubic symmetry such as silicon, the optical phonon modes, under the influence of strain, can be described by the direction of movement of the atoms in a lattice:
\begin{equation} \label{eqn:6.1:force_constant}
    \sum_{j} K_{i j} u_j = \omega ^2 u_i, \hspace{1cm} i, j = 1, 2, 3 \hspace{0.1cm},
\end{equation}
where $K_{i j}$ are the elements of the force constant matrix, $\vec{u}$ is the direction of movement of the atoms, and $\omega$ is the frequency of vibration. If the eigenvectors are along the $\hat{x}, \hat{y}, \hat{z}$ directions (as the optical modes are for silicon), we can expand the force constant matrix in powers of the strain, $\epsilon_{kl}$,
\begin{equation}
    K_{i j} = \sum_{kl} \epsilon_{kl} K_{klij}^{(\epsilon)} + K_{ij}^{(0)}\hspace{0.1cm}, \hspace{1cm} K_{ij}^{(0)}=\omega_0^2 \delta_{ij} \hspace{0.1cm},
\end{equation}
written in Einstein notation, up to first order. The new tensor $K_{klij}^{(\epsilon)}$ and the elastic tensor are both fourth-order tensors and describe two different components of elasticity of a material system. $K_{klij}^{(\epsilon)}$ is the so-called \textit{Phonon Deformation Potential} (PDP) tensor, a material constant, whose values can be computed using \textit{ab-initio}. The PDP is defined as the expectation value of the electron–phonon interaction potential.\cite{Schattke_PISS_2003} Due to the symmetry of bulk silicon, the PDPs are known and given by three unique values ($p,q,r$), in the same way that the elastic tensor is described by three values. They reduce the tensor to the same symmetry as the elastic one, giving,
\begin{equation} \label{eqn:6.1:pqr_matrix}
    \begin{pmatrix}
        K_{11}\\
        K_{22}\\
        K_{33}\\
        K_{23}\\
        K_{31}\\
        K_{12}
    \end{pmatrix}=
    \begin{pmatrix}
        p & q & q & 0 & 0 & 0\\
        q & p & q & 0 & 0 & 0\\
        q & q & p & 0 & 0 & 0\\
        0 & 0 & 0 & r & 0 & 0\\
        0 & 0 & 0 & 0 & r & 0\\
        0 & 0 & 0 & 0 & 0 & r\\
    \end{pmatrix}
    \begin{pmatrix}
        \epsilon_{11}\\
        \epsilon_{22}\\
        \epsilon_{33}\\
        2\epsilon_{23}\\
        2\epsilon_{31}\\
        2\epsilon_{12}
    \end{pmatrix}+
    \begin{pmatrix}
        \omega_0^2\\
        \omega_0^2\\
        \omega_0^2\\
        0\\
        0\\
        0
    \end{pmatrix} \hspace{0.1cm}.
\end{equation}

With the consideration of \autoref{eqn:6.1:pqr_matrix}, \autoref{eqn:6.1:force_constant} can be reduced to the secular equation,
\begin{widetext} \label{eqn:6.2:secular}
\[
    \begin{vmatrix}
        p \epsilon_{11} + q(\epsilon_{22} + \epsilon_{33}) -\lambda & 2r\epsilon_{12} & 2 r \epsilon_{31}\\
        2 r \epsilon_{12} & p \epsilon_{22} + q(\epsilon_{11} + \epsilon_{33}) -\lambda & 2 r \epsilon_{23}\\
        2 r \epsilon_{31} &  2 r \epsilon_{23} & p\epsilon_{33} + q(\epsilon_{11} + \epsilon_{22}) -\lambda\\
    \end{vmatrix} = 0,
\]
\end{widetext}
where the eigenvalues are described by $\lambda = \omega^2-\omega_0^2$. We now have a relationship that relates the strain, the PDPs, and the shift in phonon frequency from its natural state.

In its current state, the previous secular equation has no relation to the elastic tensor of silicon or the stress in a system. In order to link these quantities, recall that the stress tensor can be found from the strain of a system by,
\begin{equation}
    \vec{\sigma} = \textbf{C} \vec{\epsilon} \hspace{0.1cm},
    \label{eqn:stress_strain_voigt}
\end{equation}
or rewritten so that the strain can be replaced in the constitutive equations as
\begin{equation}
    \vec{\epsilon} = \textbf{S} \vec{\sigma} \hspace{0.1cm},
    \label{eqn:strain_stress_voigt}
\end{equation}
where $\vec{\sigma}$ is the Cauchy stress and $\textbf{S}$ is the fourth-order symmetric \textit{compliance} tensor such that it is the inverse of the elastic tensor ($\textbf{S}=\textbf{C}^{-1}$). Assuming only unixaxial stress is non-zero, $\sigma_{11}, \sigma_{22}, \sigma_{33} \neq 0$, then the strains can be written as,

\begin{equation}
\begin{aligned}
    \epsilon_{11} &= S_{11} \sigma_{11} + S_{12} \sigma_{22} +S_{13} \sigma_{33}\\
    \epsilon_{22} &= S_{21} \sigma_{11} + S_{22} \sigma_{22} +S_{23} \sigma_{33}\\
    \epsilon_{33} &= S_{31} \sigma_{11} + S_{32} \sigma_{22} +S_{33} \sigma_{33}\\
     0 &= \epsilon_{11}, \epsilon_{22}, \epsilon_{33}  \hspace{0.1cm}.
\end{aligned}
\end{equation}
Where all other strains are zero due to the symmetry of elastic tensor and due to the uniaxial stress condition, resulting in a diagonal determinant, giving three distinct solutions to the root after substitution,

\begin{equation}
    \begin{aligned}
        \lambda_1 &= p \left(S_{11} \sigma_{11} + S_{12} \sigma_{22} +S_{13} \sigma_{33} \right) \\ &+ q [ \left(S_{21} \sigma_{11} + S_{22} \sigma_{22} +S_{23} \sigma_{33} \right) \\ &+ \left(S_{31}\sigma_{11} + S_{32} \sigma_{22} +S_{33} \sigma_{33}\right) ] \hspace{0.1cm},
    \end{aligned}
\end{equation}

\begin{equation}
    \begin{aligned}
        \lambda_2 &= p \left( S_{21} \sigma_{11} + S_{22} \sigma_{22} +S_{23} \sigma_{33} \right) \\ &+ q [ \left( S_{11} \sigma_{11} + S_{12} \sigma_{22} +S_{13} \sigma_{33} \right) \\ &+ \left( S_{31} \sigma_{11} + S_{32} \sigma_{22} +S_{33} \sigma_{33} \right) ] \hspace{0.1cm},
    \end{aligned}
\end{equation}

\begin{equation}
    \begin{aligned}\label{eqn:4.1:lambda3_constitutive}
        \lambda_3 &= p \left( S_{31} \sigma_{11} + S_{32} \sigma_{22} +S_{33} \sigma_{33} \right) \\ &+ q [ \left( S_{11} \sigma_{11} + S_{12} \sigma_{22} +S_{13} \sigma_{33} \right) \\ &+ \left( S_{21} \sigma_{11} + S_{22} \sigma_{22} +S_{23} \sigma_{33} \right) ]  \hspace{0.1cm}.
    \end{aligned}
\end{equation}

The result is one which suggests that the phonon frequency shifts detected under strain depend on both the stiffness of the system, the stress applied, and the natural frequency of the system. Depending on the direction and frequency of incident light, the excited phonon mode will differ. In the case of [001] incident polarized light, or light polarized along the Cartesian-$\hat{z}$, then only the third longitudinal optical peak will be observed, leaving only the $\lambda_3$ equation. From here, depending on the stress in the system, the result can be further simplified. Current characterization using Raman spectroscopy relies on these material constants, in particular the bulk values, to measure the stress in complex semiconductor devices, but the stiffness of ultra-thin free-standing films or stacks of materials is not the same as in bulk.

\section{Appendix:Computation of Ultra-Thin Elastic Tensors}

The procedure for calculating ultra-thin crystalline elastic tensors can be found in Ref.~\cite{Pashartis_ASS_2022}, where it was performed on Ru (001) films. The only difference between Ru and Si is the requirement of a larger unit cell to properly accommodate the surface reconstruction in the 2x1 [001] direction. For completeness we have plotted the elastic tensor as an evolution of thickness, with with the edge-to-edge thickness and the Hartree corrected scheme in \autoref{fig:Si_001_dzvp_corrected_vs_non_corrected_v2}. The latter takes into account the extent of the Hartree potential extending past the surface compared to the bulk-type system, see reference. It is interesting to note that the proper trend for decreasing thickness is only observed with the corrected thickness schemed. The methodology was developed such that a consistent ratio can be defined over the same surface, regardless of thickness. In this case we determined a $I/I_0$ ratio of 0.2 (see \autoref{fig:Si_1nm_unrelaxed_slab_determined_ratio}). As a proof of concept and to demonstrate that our tensor is correctly computed, we compare the tensor computed Young's Modulus (which takes into account various components of the tensor) with DFT computed literature. 

\makefigure{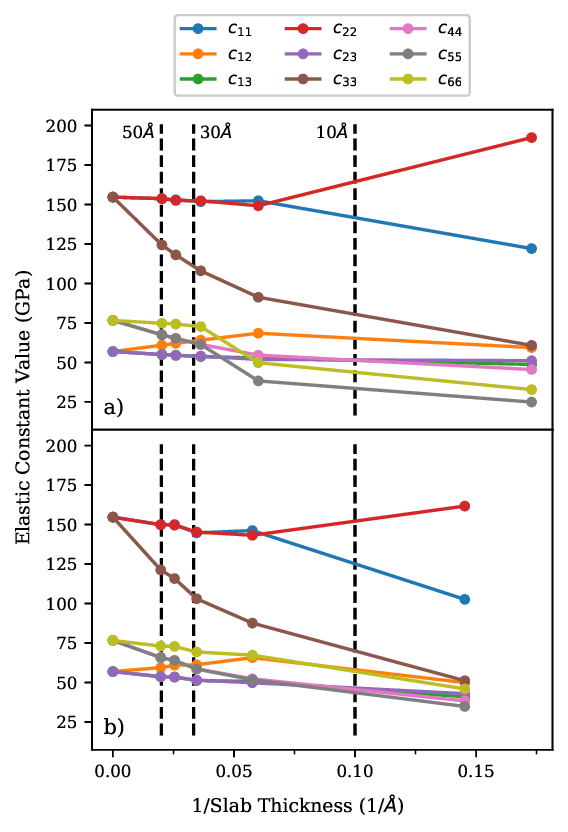}{Elasticity of ultra-thin Si $2\times1$ (001), where a) is the uncorrected stress tensor edge-to-edge thickness and b) is the corrected version using the Hartree scheme. The different colours correspond to the different elastic tensor components, $c_{13}$ is typically underneath $c_{12}$.}

\makefigure{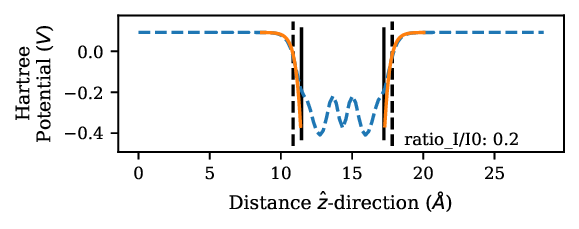}{Hartree potential fitted curve of a 6~\r{A} edge-to-edge thick slab. The blue line is the Hartree potential, the orange line is the fitted tail to the edge of the surface, the solid black one is the atomic edge of the unreconstructed surface, and the dotted black line is the height of the equivalent bulk system.}

\makefigure{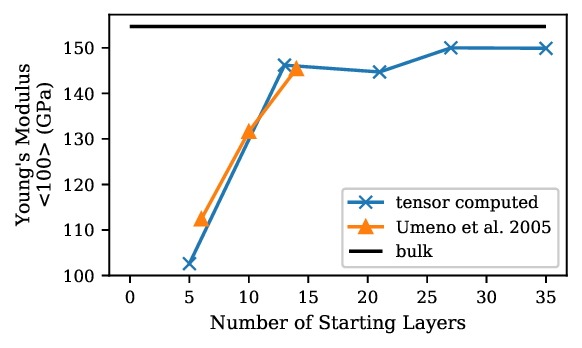}{Young's modulus computed in the [100] direction of Si $2\times1$ (001) surfaces comparing the elastic tensor derived value and with strained results of Ref.~\cite{Umeno_PRB_2005}}

Elastic tensor calculations were performed using the first-principle \texttt{CP2K} simulation package \cite{Borstnik_PC_2014, Hutter_WIRCMS_2014, Guidon_TJOCP_2008, Guidon_JOCTAC_2009, Guidon_JOCTAC_2010, Kuehne_TJOCP_2020}, using 3D periodic boundary conditions. A Fermi Dirac distribution for the occupancy of the valence band structure has been imposed with an electronic temperature of 1000~K. The Goedecker-Teter-Hutter \texttt{CP2K} pseudopotential library \cite{Goedecker_PRB_1996, Hartwigsen_PRB_1998, Krack_TCA_2005}, was used and combined with the Perdew–Burke-Ernzerhof (PBE) \cite{Perdew_PRL_1996} exchange correlation functional. DZVP basis sets were used from the \texttt{CP2K} standard library.\cite{VandeVondele_TJOCP_2007} The simulations were converged with respect to the elastic tensor components for the bulk conventional structure to achieve a precision of 10~GPa. The bulk lattice parameters and elastic components were found to be converged with a k-point Monkhorst pack grid mesh of $7\times7\times7$, a kinetic energy cutoff of 200~Ry and a relative energy cutoff of 50~Ry (for the Gaussian grid). The maximum force cutoff used for unit cell relaxation was 1E-4~bohr$^{-1}$~Ha. Optimization was performed for the standard cubic unit cell of Si.

\end{document}